\documentclass[twocolumn, showpacs]{revtex4-1}
\usepackage{amsmath}
\usepackage{amsfonts}
\usepackage{amssymb}
\usepackage{amscd}
\usepackage{cancel}
\usepackage{ntheorem}
\usepackage[american]{babel}
\usepackage[T1]{fontenc}
\usepackage{textcomp}
\usepackage[dvips]{rotating}
\usepackage{fancybox}
\usepackage{eurosym}
\usepackage[utf8]{inputenc}
\usepackage{epsfig}
\usepackage[pdftex]{color}
\usepackage{graphicx}
\usepackage{palatino}
\usepackage{array}

\newcommand{\avg}[1]{\left< #1 \right>} 
 
\newcommand{\ket}[1]{\left| #1 \right>} 

\let\baraccent=\= 
\renewcommand{\=}[1]{\stackrel{#1}{=}} 

\renewcommand{\b}[1]{{\bf #1}}
\renewcommand{\i}{\mathrm{i}}
\def\beq#1\eeq{\begin{equation}\begin{split}#1\end{split}\end{equation}}

\newcommand{\bmat}{\begin{pmatrix}}
\newcommand{\emat}{\end{pmatrix}}
\newcommand{\bdet}{\begin{dmatrix}}
\newcommand{\edet}{\end{dmatrix}}

\newcommand{\Lra}{\Leftrightarrow}

\newcommand{\tn}[1]{\textnormal{#1}}

\begin{document}
\title[Quantum many--body systems with the GKBA]{Non--equilibrium Green's function approach to inhomogeneous quantum many--body systems using the Generalized Kadanoff Baym Ansatz}
\author{S Hermanns}
\email{hermanns@theo-physik.uni-kiel.de}
\author{K Balzer}
\author{M Bonitz}
\affiliation{$^1$ Institut f\"ur Theoretische Physik und Astrophysik, Christian-Albrechts-Universit\"at Kiel, Leibnizstrasse 15, 24098 Kiel, Germany}
\begin{abstract}
In non--equilibrium Green's function calculations the use of the Generalized Kadanoff--Baym Ansatz (GKBA) allows for a simple approximate reconstruction of the two--time Green's function from its time--diagonal value.
With this a drastic reduction of the computational needs is achieved in time--dependent calculations, making longer time propagation possible 
and more complex systems accessible. 
This paper gives credit to the GKBA that was introduced 25 years ago. After a detailed derivation of the GKBA, we recall its application to homogeneous systems and show how to extend it to strongly correlated, inhomogeneous systems.
As a proof of concept, we present results for a 2--electron quantum well, where the correct treatment of the correlated electron dynamics is crucial for the correct description of the equilibrium and dynamic properties.
 
\end{abstract}
\pacs{05.10.-a, 05.30.-d, 71.10.-w}
\maketitle
\section{Introduction}

For the time--dependent description of non--equilibrium processes the method of non--equilibrium Green's function (NEGF) has been widely used, since it allows for self--consistent treatment of electron--electron correlations, non--perturbative inclusion of external fields and systematic approximations via Feynman diagrams. The central quantity is the one--particle, two--time function, $G(t,t')$, the time--evolution of which is governed by the Keldysh--Kadanoff--Baym--equations~\cite{KadanoffBaym1962}.
To numerically solve these equations a self--energy is introduced, which can be determined by many--body perturbation theory and which leads to a closed equation for $G$. Still, as $G$ inherently depends on two times $t$ and $t'$, the time propagation is numerically demanding and the memory and CPU time needs scale quadratically with the propagation length,  see e.g. \cite{balzer_efficient_2010}.    
 
This restriction can be drastically alleviated by the introduction of a further approximation, the {\em generalized Kadanoff--Baym ansatz} (GKBA), which was introduced by Lipavsky, Spicka and Velicky some 25 years ago~\cite{lipavsky_generalized_1986}. With the GKBA, for each time propagation step, the two--time Green's function is reconstructed from its time--diagonal value: $G(t,t')=F_{\tn{GKBA}}[G(t=t')]$.
As a great advantage it reduces the amount of needed memory to a linear scaling with propagation length, since for the determination of $G(T,T')$ for time--arguments $T' \leq T$ only the knowledge of $G(t=t')$ for all $t \leq T$ is sufficient.
This simplification has made numerous applications for spatially homogeneous systems possible. Here we demonstrate that the GKBA may be equally successful in computing the behaviour of {\em finite inhomogeneous systems}.

The paper is organized as follows: After a short recollection of the basics of the NEGF formalism we, in detail, derive and list the properties of the GKBA. Thereafter we give a brief overview about its application to  homogeneous systems in different fields of physics.
In a third part we extend the GKBA to inhomogeneous systems, using the technique of adiabatic switching (AS)~\cite{rios_time-dependent_2008} to correctly obtain the associated correlated initial state.
Finally the applicability of the GKBA to the spctrum of a two--electron quantum well is tested at different coupling strengths.       
\section{Theory}
\subsection{Non--equilibrium Green's Functions (NEGF)}
To describe correlation effects and excitations in quantum many--particle systems we chose the NEGF approach, as it allows for a systematical inclusion of correlations by diagrammatic expansions. In contrast to density matrix based schemes, the Green's function method additionally easily offers direct access to dynamical spectral information as well as particle removal and addition energies.    
The main quantity is the one--particle Green's function, defined as (we set $\hbar~\equiv~1$)
\begin{align}
G(t,t') = -\i\avg{\mathcal{T}_\mathcal{C}\left[\Psi(t)\Psi^\dagger(t')\right]},
\end{align}
where the brackets denote thermodynamical averaging and $\mathcal{T}_\mathcal{C}$ is the time ordering operator on the Schwinger--Keldysh contour $\mathcal{C}$~\cite{KeldyshContour}, on which $t$ and $t'$ are defined. $\Psi^{(\dagger)}$ denotes a one--particle annihilation (creation) operator in a one--particle basis in second quantization.
The equations of motion for $G$ are the Keldysh--Kadanoff--Baym equations (KBE)
\begin{align}
\label{eq:KBE}
\left[i\partial_{t}-h(t)\right]G(t,t')&=\delta(t-t') \\
&\quad+\int_\mathcal{C} d\bar t\,W(t,\bar t)\,G^{(2)}(t\bar t;\,t' \bar t^+) \,,\nonumber\\
\left[-i\partial_{t'}-h(t')\right]G(t,t')&=\delta(t-t')\nonumber\\
&\quad+ \int_\mathcal{C} d\bar t\,W(t',\bar t)\,G^{(2)}(t\bar t;\,t' \bar t^+),\nonumber
\end{align}
where $h$ denotes the one--particle Hamiltonian,\\ $G^{(2)}(tt';\,t'_1t'_2)=-\avg{\mathcal{T}_\mathcal{C}\left[\Psi(t)\Psi(t')\Psi^\dagger(t'_2)\Psi^\dagger(t'_1)\right]}$ is the two--particle Green's function and $W$ is an arbitrary 
interaction potential.
The KBE are the first equations of the Martin--Schwinger Hierarchy~\cite{martin_theory_1959}, which describes the coupling of the evolution of the one--particle Green's function to the two--particle Green's function, which itself is coupled to the three--particle Green's function by a similar equation. To decouple this hierarchy and to make the KBE numerically tractable, a self--energy $\Sigma(t,t')=\Sigma[G(t,t')]$ is introduced. This self--energy can be found from a diagrammatic expansion in terms of Feynman diagrams, where only some classes of diagrams are chosen according to the properties of the examined system.
With this the KBE attain a formally closed form:  
\begin{align}
\label{eq:KBESigma}
\left[\i\partial_{t}-h(t)\right]G(t,t')&= \delta(t-t')+ \int_\mathcal{C} d\bar t\,\Sigma[G](t,\bar t)G(\bar t,t')\,, \nonumber\\\nonumber\\
\left[-\i\partial_{t'}-h(t')\right]G(t,t')&= \delta(t-t') + \int_\mathcal{C} d\bar t\,G(t,\bar t)\Sigma[G](\bar t,t')\,.\nonumber\\
\end{align}

\subsection{Reconstruction Problem}
\subsubsection{Keldysh representation and Dyson equation}
The two--time structure of the time contour $\mathcal{C}$ suggests the use of a matrix representation for $\mathbf{G}$ according to the different time orderings. There exist different representations, that are connected by a Keldysh rotation~\cite{KeldyshKamenev}. Here, we use the set involving less--, retarded and advanced Green's functions ($G^<, G^R$ and $G^A$) according to Langreth and Willkins~\cite{LWR},
\begin{equation}
\setlength{\extrarowheight}{0.5ex}
\mathbf{G}=\left(\begin{array}{cc}
G^R & G^< \\ 0 & G^A
\end{array} \right)\,.
\end{equation} 
Note, that this representation implies a two--time dependence of each component and is overcomplete as the conjugation relation $\left[G^R(t,t')\right]^* = G^A(t,t')$ holds.
To simplify the notation we will make use of the greater Green's function $G^>$, which relates to the other components as:
\begin{equation}
G^>(t,t')=G^R(t,t')-G^A(t,t')+G^<(t,t')\,.
\end{equation}
With these definitions we arrive at a formal solution for $\mathbf{G}$ by time integration of Eq.~(\ref
{eq:KBESigma}) yielding the nonequilibrium version of the Dyson equation,
\begin{equation}
\label{eq:Dyson1}
\mathbf{G} = \mathbf{G}_0+\mathbf{G}_0\mathbf{\Sigma} \mathbf{G}\,.
\end{equation}
Here $\b{G}_0$ denotes the non--interacting or Hartree--Fock (HF) Green's function, whose inverse is given by $\b{G}_0^{-1}$:
\begin{equation}
\setlength{\extrarowheight}{0.5ex}
\mathbf{G}_0^{-1}={G}_0^{-1}\cdot\mathbf{1}=\left(\begin{array}{cc} {G}_0^{-1} & 0 \\ 0 & {G}_0^{-1}\end{array}\right) , 
\end{equation}
where $(G_0)^{-1}$, the common inverse of the components $G^R$ and $G^A$, reads
\begin{align}
\label{eq:G01}
(G_0)^{-1}(t,t')=\delta(t,t')\left[\i\partial_t-h(t)\,\right].
\end{align}
The matrix multiplication in Eq.~(\ref
{eq:Dyson1}) is to be understood as also including a time integration on the contour over intermediate time coordinates, so that, e.g.
\begin{equation}
\label{eq:DefContInt}
\left(\mathbf{G}_0\mathbf{\Sigma} \mathbf{G}\right)(t',t)=\int_{\mathcal{C}}d\bar t\,d\bar{\bar t}\,\mathbf{G}_0(t,\bar{t})\mathbf{\Sigma}(\bar{t},\bar{\bar{t}}) \mathbf{G}(\bar{\bar{t}},t')\,.
\end{equation}
\subsubsection{Equation of motion for $G^{<}$ in terms of the density matrix}
Following V. Spicka et al.~\cite{spicka_long_2005}, an intermediate step towards the GKBA is to express the equation of motion of $G^<$ in terms of the density matrix $\rho(t)=-iG^<(t,t)$. To start with, we provide some useful relations between $G^{R/</A}$ and the respective self--energies $\Sigma^{R/</A}$.

By right--multiplication of Eq.~(\ref
{eq:Dyson1}) with $\mathbf{G}^{-1}$ and left--multiplication with $(\mathbf{G}_0)^{-1}$ we attain, taking the retarded/advanced component: 
\begin{equation}
\label{eq:Dyson2}
\left(G_0\right)^{-1} = G^{-1,R/A} + \Sigma^{R/A}.
\end{equation}
Now taking the less--component of Eq.~(\ref{eq:Dyson1}), after left--multiplication with $(\b{G}_0)^{-1}$, we find
\begin{equation}
\left((\mathbf{G}_0)^{-1}\mathbf{G} \right)^< = \mathbf{1}^< + \left(\mathbf{\Sigma} \mathbf{G}\right)^< = \left(\mathbf{\Sigma} \mathbf{G}\right)^<.
\end{equation}
Using the Langreth--Wilkins Rules~\cite{LWR} it follows:
\begin{equation}
\label{eq:Dyson3}
\left(G_0\right)^{-1,R}G^<+\left(G_0\right)^{-1,<}G^A = \Sigma^R G^<+\Sigma^< G^A.
\end{equation}
Note, that the multiplication is to be understood in the same manner as in Eq.~(\ref
{eq:DefContInt}), including contour time integration. Since $\left(G_0\right)^{-1,<} \equiv 0$, Eq.~(\ref
{eq:Dyson3}) simplifies to:
\begin{equation}
\label{eq:G0}
\left(G_0\right)^{-1}G^< = \Sigma^R G^<+\Sigma^< G^A,
\end{equation}
and use of Eq.~(\ref{eq:Dyson2}) yields:
\begin{equation}
\left(G^{-1,R} + \Sigma^R\right) G^< = \Sigma^R G^<+\Sigma^< G^A.
\end{equation}
Analogously we find the conjugate equation, resulting in two final differential equations:
\begin{align}
\label{eq:DiffG}
G^{-1,R}G^< &= \Sigma^< G^A, \\
G^< G^{-1,A} &= G^R \Sigma^<.
\end{align}
Now it is convenient to also split $G^<$ into two parts corresponding to the time arguments $t>t'$ and $t\leq t'$:
\begin{align}
\label{eq:G_def}
G^<(t,t') &= G^<_R(t,t')-G^<_A(t,t'),\\
G^<_R(t,t') &= \Theta(t-t') G^<(t,t'), \\
G^<_A(t,t') &= - \Theta(t'-t)G^<(t,t').
\end{align}
This allows us to separately derive an equation for $G^<_R$ and $G^<_A$ from which Eq.~(\ref
{eq:G_def}) allows us to recover the equation for $G^<$. For $G^<_R$, one calculates:
\begin{align}
\label{eq:Grless}
&\left[G^{-1,R}G^<_R\right](t,t') \\
&=\left[\left((G_0)^{-1}-\Sigma^R\right)G^<_R\right](t,t')\nonumber\\
&= \int_\mathcal{C}d\bar t\,\left[\delta(t,\bar t)\left(\i\partial_{\bar t}-h(\bar t)\,\right)\Theta(\bar t-t')G^<(\bar t,t')\right] \nonumber\\
&\quad-\int_\mathcal{C}d\bar t\,\Sigma^R(t,\bar t) G^<_R(\bar t,t') \nonumber\\
&= \int_\mathcal{C}d\bar t\,\left[\delta(t,\bar t)\left(\i\delta(\bar t, t')G^<(\bar t,t')+\i\Theta(\bar t-t')\partial_{\bar t} G^<(\bar t, t')\right.\right. \nonumber\\
&\quad\left.\left.-h(\bar t)\Theta(\bar t-t')G^<(\bar t,t')\right)\right] -\int_\mathcal{C}d\bar t\,\Sigma^R(t,\bar t) G^<_R(\bar t,t') \nonumber\\
&= \delta(t-t')\,\i G^<(t,t') \nonumber\\
&\quad+\int_\mathcal{C}d\bar t\,\delta(t,\bar t)\Theta(\bar t-t')\left(\i\partial_{\bar t}-h(\bar t)\right)G^<(\bar t, t')\nonumber\\
&\quad-\int_\mathcal{C}d\bar t\,\Sigma^R(t,\bar t) G^<_R(\bar t,t') \nonumber\\
&= \delta(t-t')\,\i G^<(t,t')\nonumber\\
&\quad+\Theta(t-t')\int_\mathcal{C}d\bar t\,\Theta(\bar t-t')(G_0)^{-1}(t,\bar t)G^<(\bar t, t')\nonumber\\
&\quad-\int_\mathcal{C}d\bar t\,\Sigma^R(t,\bar t) G^<_R(\bar t,t') \nonumber\\
&= \delta(t-t')\,\i G^<(t,t')\nonumber\\ 
&\quad+\Theta(t-t')\left(\int_\mathcal{C}d\bar t\,\Theta(\bar t-t')G^{-1,R}(t,\bar t)G^<(\bar t, t')\right.\nonumber\\
&\quad\left.+\int_\mathcal{C}d\bar t\,\Sigma^R(t,\bar t)G^<_R(\bar t,t')-\int_\mathcal{C}d\bar t\,\Sigma^R(t,\bar t) G^<_R(\bar t,t')\right) \,,\nonumber
\end{align}
where the inclusion of the last term under the $\Theta$--function is justified, since the contour product of two retarded functions is again a retarded function.
The two last terms cancel and one finds, employing Eq.~(\ref
{eq:DiffG}),
\begin{align}
&\delta(t-t')\,\i G^<(t,t')\\ 
&\quad+\Theta(t-t')\int_\mathcal{C}d\bar t\,\Theta(\bar t-t')G^{-1,R}(t,\bar t)G^<(\bar t, t')\nonumber\\
=\, &\delta(t-t')\,\i G^<(t,t')\nonumber\\ 
&\quad+\Theta(t-t')\int_\mathcal{C}d\bar t\,\Sigma^<(t,\bar t)G^A(\bar t, t')\nonumber\\
&\quad-\Theta(t'-t)\int_\mathcal{C}d\bar t\,\Theta(t'-\bar t)G^{-1,R}(t,\bar t)G^<(\bar t, t')\,.\nonumber
\end{align}
Here, the last term compensates the step function in the second term. With Eq.~(\ref{eq:Dyson2}) and, noting that $$\int_\mathcal{C}d\bar t\,\Theta(t'-\bar t)\left(G_0\right)^{-1}(t,\bar t)G^<(\bar t,t')\equiv 0\,,$$ it follows 
\begin{align}
&\delta(t-t')\,\i G^<(t,t')\\ 
&\quad+\Theta(t-t')\int_\mathcal{C}d\bar t\,\Sigma^<(t,\bar t)G^A(\bar t, t')\nonumber\\
&\quad-\Theta(t'-t)\int_\mathcal{C}d\bar t\,\Theta(t'-\bar t)G^{-1,R}(t,\bar t)G^<(\bar t, t')\nonumber\\
&=\,\delta(t-t')\,\i G^<(t,t')\nonumber\\ 
&\quad+\Theta(t-t')\int_\mathcal{C}d\bar t\,\Theta(t'-\bar t)\Sigma^<(t,\bar t)G^A(\bar t, t')\nonumber\\
&\quad+\Theta(t-t')\int_\mathcal{C}d\bar t\,\Theta(t'-\bar t)\Sigma^R(t,\bar t)G^<(\bar t, t')\,,\nonumber
\end{align}
where, in the second term, the fact was used, that $G^A$ includes a step function by definition. Left--multiplying by $G^R$ and explicitly writing out the integrals, one arrives at the equation for $G^<_R(t,t')=G^<(t,t')$, valid for times $t>t'>t_0$:
\begin{align}
\label{eq:G_R_less}
G^<(t,t') &= -G^R(t,t')\rho(t')\\
&+ \int_{t'}^{t}d\bar{\bar t}\,\int_{t_0}^{t'}d \bar t\, G^R(t,\bar{\bar t})\Sigma^<(\bar {\bar t}, \bar t)G^A(\bar t, t')\nonumber\\ &+\int_{t'}^{t}d\bar{\bar t}\,\int_{t_0}^{t'}d \bar t\,G^R(t,\bar {\bar t})\Sigma^R(\bar {\bar t}, \bar t)G^<(\bar t, t')\,.\nonumber 
\end{align}
In a similar manner one can derive the equation for $G^<(t,t')=-G^<_A(t,t')$ in the time domain $t_0<t<t'$, which reads:
\begin{align}
\label{eq:G_A_less}
G^<(t,t') &= \rho(t)G^A(t,t')\\
&+ \int_{t'}^{t}d\bar{\bar t}\,\int_{t_0}^{t'}d \bar t\, G^R(t,\bar{\bar t})\Sigma^<(\bar {\bar t}, \bar t)G^A(\bar t, t')\nonumber\\ &+\int_{t'}^{t}d\bar{\bar t}\,\int_{t_0}^{t'}d \bar t\,G^<(t,\bar {\bar t})\Sigma^A(\bar {\bar t}, \bar t)G^A(\bar t, t')\,, \nonumber
\end{align}
Note, that by exchanging ($<\Lra>$) and replacing the density matrix $\rho=:f^<$ by $f^>=1-f^<$ in equations (\ref{eq:G_R_less}) and (\ref{eq:G_A_less}), the analogous expression for $G^>$ is easily obtained.
\subsection{The Generalized Kadanoff--Baym Ansatz} 
Combining equations (\ref{eq:G_R_less}) and (\ref{eq:G_A_less}), retaining only the terms without the integrals, the Generalized~ Kadanoff--Baym~Ansatz~\cite{lipavsky_generalized_1986} is recovered:
\begin{align}
\label{eq:GKBA}
G^\gtrless_{{\tn{GKBA}}}(t,t')=-G^R(t,t')f^\gtrless(t')+f^\gtrless(t)G^A(t,t').
\end{align}
With this equation a means for the reconstruction of the off--diagonal Green's function from the density matrix is found, which obeys particle number conservation, has a per se causal structure and does not depend on assumptions about near equilibrium, for a recent discussion, see eg. Ref.~\cite{spicka_long_2005}. 
\subsubsection{Choice of propagators $G^{R/A}$}
When looking at Eq.~(\ref{eq:GKBA}), it should be noted, that the GKBA is only formally closed in terms of $\rho$, since it depends on the knowledge of the retarded (advanced) propagators $G^R\,(G^A)$, which themselves obey non--Markovian two--time equations of motion of similar complexity. This can be overcome by the use of propagators, which obey a Markovian evolution.
In this paper we choose Hartree--Fock propagators, as they incorporate the interaction at mean--field level in contrast to, e.g., ideal propagators. The HF propagators $G^{R/A}_{HF}$ are given by:
\begin{equation}
\label{eq:propagators}
G^{R/A}_{HF}(t,t')=\mp\i\Theta[\pm(t-t')]\exp\left(-\i\int_{t'}^td \bar t\,H(\bar t)\right),
\end{equation}
where $H$ denotes the mean--field HF Hamiltonian, which is governed by the time--dependent density matrix.
\subsubsection{Choice of self--energy}
To show the advantages of the GKBA we exemplarily apply the second order Born approximation, leading to a self--energy $\Sigma_{2B}(t,t')$, which is given by
\begin{align}
\Sigma_{2B}^<(t,t')=&\,\delta(t-t')\Sigma_{HF}^<(t)\nonumber\\
&+G^<(t,t')W(t)W(t')G^>(t',t)
G^<(t,t')\nonumber\\
&-G^<(t,t')W(t)G^>(t',t)
G^<(t,t')W(t')\,.
\end{align}
The evaluation of the collision integral $I=\int_\mathcal{C}\mathbf{\Sigma G}$, the right hand side of the KB equations Eq.~(\ref{eq:KBESigma}), is twofold simplified by the GKBA as we can use the reconstructed $G(t,t')$ in $I$ as well as in $\Sigma$.   
For instance, the less--part of $\Sigma_{2B}$ for $t>t'$ now reads:
\begin{align}
\Sigma_{2B}^<(t,t')=&\,\delta(t-t')\Sigma_{HF}^<(t)\nonumber\\
&+G^R(t,t')f^<(t') \nonumber\\
&\left[W(t)W(t')f^>(t')G^A(t',t)G^R(t,t')f^<(t')
\nonumber\right.\\
&-\left.W(t)f^>(t')G^A(t',t)
G^R(t,t')f^<(t')W(t')\right]\,.
\end{align}
So only the single--time quantity $f^\gtrless$ has to be stored in memory, as the HF propagators $G^{R/A}(t,t')$ can be computed each time step, and the demand scales linearly with propagation length. That way, in total three approximations were taken: \begin{itemize}
\item [i.] The self--energy was introduced to decouple the Martin--Schwinger hierarchy.
\item [ii.] The two--time Green's function was reconstructed within the GKBA according to Eq.~(\ref
{eq:GKBA}), neglecting the integral terms in equations (\ref{eq:G_R_less}) and (\ref{eq:G_A_less}). 
\item [iii.] The propagators $G^{R/A}$ were approximated by HF propagators, which can be explicitly calculated for each ($t,t'$).
\end{itemize}
\section{Application to homogeneous systems -- Overview}\label{sec:Hom}

The GKBA has been applied to spatially homogeneous charged particle systems already in the mid 1990s and has allowed for many numerical studies of ultrafast carrier relaxation. Electron--phonon scattering in semiconductors was investigated e.g. in Refs.~\cite{gartner_prb99,gartner_prb06}. Electron--electron in plasmas and semiconductors as well as
tests of the GKBA against full two--time calculations were performed in Refs.~\cite{bonitz_jpcm96,koehler_pre96,kwong_pss98}. The use of correlated spectral functions (beyond Hartree--Fock propagators) was analyzed in Ref.~\cite{bonitz_epjb}.
For an overview on the GKBA and applications to electron--hole plasmas in semiconductors, see the text book \cite{HaugJauho}.
The second type of applications was devoted to dense plasmas, in particular laser plasmas. To capture strong field effects in the Coulomb scattering process (such as harmonics generation, inverse bremsstrahlung), a gauge--invariant generalization of the GKBA was derived in Ref.~\cite{kremp_pre99} and used in \cite{bonitz_cpp99,haberland_pre01}.

Since application of the GKBA transforms the NEGF approach into a single--time theory, there should exist close connections to the purely 
single--time approach of reduced density operators (quantum BBGKY--hierarchy). In fact, this  has been studied in detail in Ref. \cite{bonitz-book}, and a one to one correspondence can be established in the limit of free or Hartree--Fock propagators. There it was also shown that the GKBA does not destroy the conserving properties of the underlying NEGF approximation as long as no appoximations to the time structure of the propagators are introduced. In particular, in the relevant case of free or HF propagators the conservation laws and sum rules are preserved \cite{bonitz_pla96}.
For a recent overview on the GKBA and its relation to quantum transport and density functional theory, see \cite{velicky_jpcs06}.
\section{Application to inhomogeneous systems}
\subsection{Initial state preparation under the GKBA} 
In order to solve the KBE Eq.~(\ref{eq:KBESigma}), one has to supply an initial value for $G(t_0,t_0)$.
For large, homogeneous systems described in Section \ref{sec:Hom}, the initial state is reached from an arbitrary state after a characteristic equilibration time.
For small, strongly correlated systems, no equilibration takes place and the preparation of the correlated initial state has to be performed differently to avoid strong artifacts, such as unphysical oscillations.
For full two-time calculations, the initial state can be found by means of the extended Matsubara--Schwinger--Keldysh time--contour (containing an additional imaginary time branch) and solution of the Dyson equation Eq.~(\ref{eq:Dyson1}) on its imaginary branch, e.g.~\cite{VetterWalecka}, using the same self--energy (for details and implementation see e.g. \cite{stan_time_2009}).
Here, however, the application of the GKBA poses a problem, since no corresponding equilibrium approximation is known so far.
This can be remedied by using adiabatic switching as discussed below.    

\begin{center}
\begin{figure}
\includegraphics[width=8.0cm]{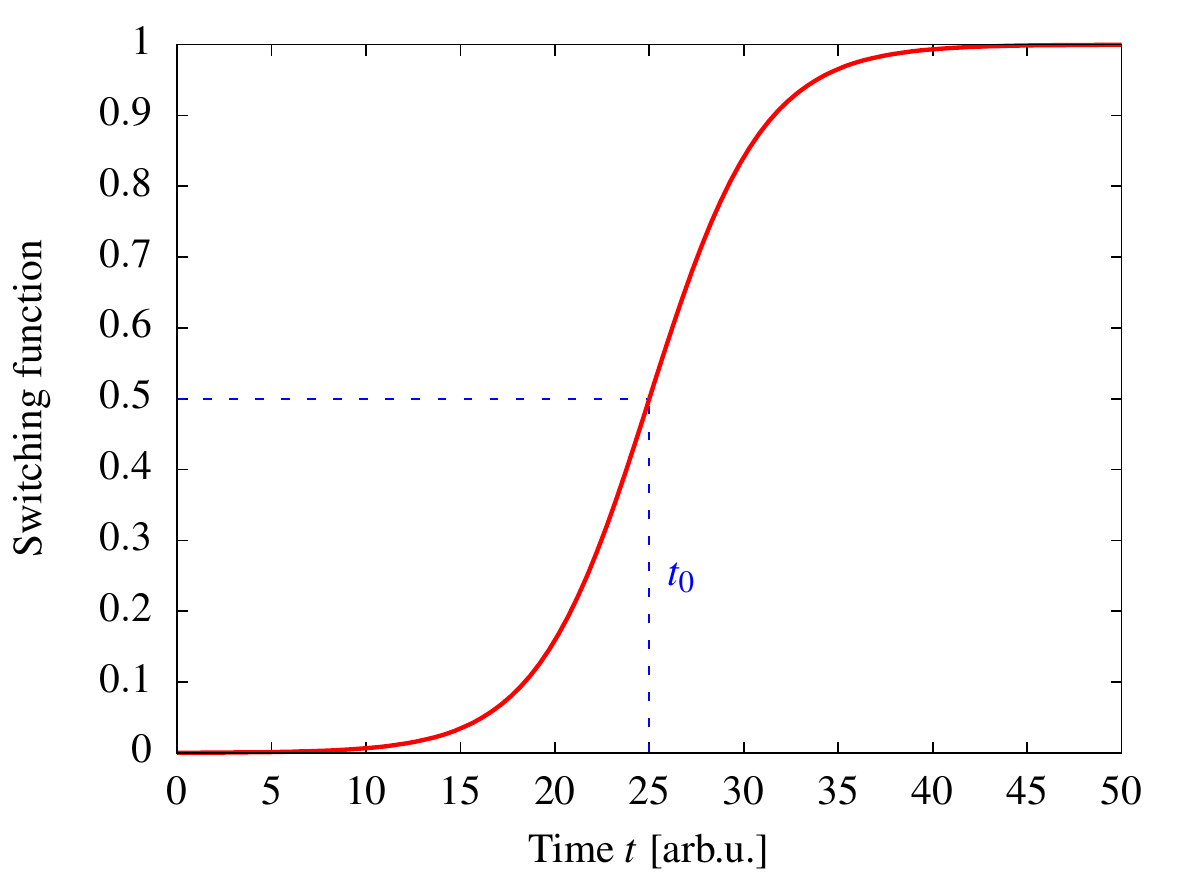}
\caption{Fermi function for adiabatic switching. The half--time is $t_0=25$ and the switching time constant is chosen to be $\tau=3$}
\label{fig:Fermifunction}
\end{figure}
\end{center}

In general, according to the Adiabatic Theorem~\cite{born_beweis_1928}, the ground--state of an interacting system can be found by taking the ideal system, for which the ground--state is assumed to be known, and adiabatically switching on the interaction.
If this procedure is performed slowly enough, the system is undergoing a transition through successive eigenstates of the respective Hamiltonians with gradually higher interaction strength. Finally, the fully interacting ground--state is reached.
Here, we use this method for generating the initial state that is consistent with the GKBA. We use a monotonic switching function $f$, which has the following properties:
\begin{align}
f(t_0)=0, \quad f(t_{f})=1,\quad 0 \leq f(t_0 < t < t_f) \leq 1 \,,
\end{align}
where $t_f$ is the end of the switching process and the start of the time--dependent calculation. It is important that, with this method, the interaction that normally is time--independent, becomes inherently time--dependent.
This especially needs to be kept in mind, when dealing with quantities involving interaction terms. For example, the HF propagators now include $W(t)$  and the self--energy contains $W(t)$ and $W(t')$. In the calculations below we use a Fermi--like switching function  $f_{t_0}^{\tau}(t)$ (see Fig. \ref{fig:Fermifunction}): 
\begin{equation}
f_{t_0}^{\tau}(t) = 1-\frac{1}{1+\exp\left(\frac{t-t_0}{\tau}\right)},
\label{eq:Fermifunction}
\end{equation}
where $f_{t_0}^{\tau}(t_0)=\frac{1}{2}$ and the smoothness of the  switching increases with the value of time constant $\tau$.

\subsection{Application to electrons in quantum wells}
To test the ability of the GKBA to describe correlation effects in an inhomogeneous system, we study two electrons in a quantum well potential.
After preparing the system in the correlated initial state via adiabatic switching the system is disturbed by a short dipole kick \cite{kwong_prl00} with sufficiently small amplitude, and the time--dependent dipole moment is computed. Fourier transformation then yields the correlated dipole excitation spectrum in linear response with the relevant vertex corrections, thereby fully preserving conservation laws and sum rules~\cite{kwong_prl00}.
All calculations are performed in the context of a FE--DVR basis, which drastically reduces the numerical complexity, for details the reader is referred to refs.~\cite{balzer_efficient_2010,balzer_time-dependent_2010}.

We consider $N=2$ electrons in a quantum well potential, which is effectively a one--dimensional system, if the lateral electronic motion is neglected. 
We assume, that the system is in singlet configuration $\ket{S,M_S}=\ket{0,0}$.
The confinement energy is given by $E_0^*=\hbar^2/(m^*L^2)$, where $m^*$ is the effective mass of the electrons within the quantum well potential of width $L$.
The 2--particle Hamiltonian in units of $E_0^*$ reads $H_2(t)=\left(\sum_{i=1}^{2}-\frac{1}{2}\frac{\partial^2}{\partial x_i^2}+f_0\cdot x_i\,\delta(t-t_0)\right)+\lambda^*\left[(x_1-x_2)^2+\kappa^2\right]^{-1/2}$, where $x_i$ are the particle positions. The first term denotes the single--particle contributions involving the potential energy and the dipole delta--kick with amplitude $f_0 \ll 1$. 

The second term describes the two--particle Coulomb interaction. A cut--off parameter $\kappa$ has been inserted to regularize the 1D--Coulomb potential, which is set to a value of 1 throughout the calculations. The relative interaction strength between the electrons is given by the dimensionless coupling parameter $\lambda^*=L/a_0^*=e^2m^*L/(4\pi\epsilon_0
\epsilon^*\hbar^2)$, where $\epsilon^*$ denotes the material's dielectric constant that enters the effective Bohr radius $a_0^*$. 
\subsection{Numerical results}
\begin{center}
\begin{figure}
\includegraphics[width=8.5cm]{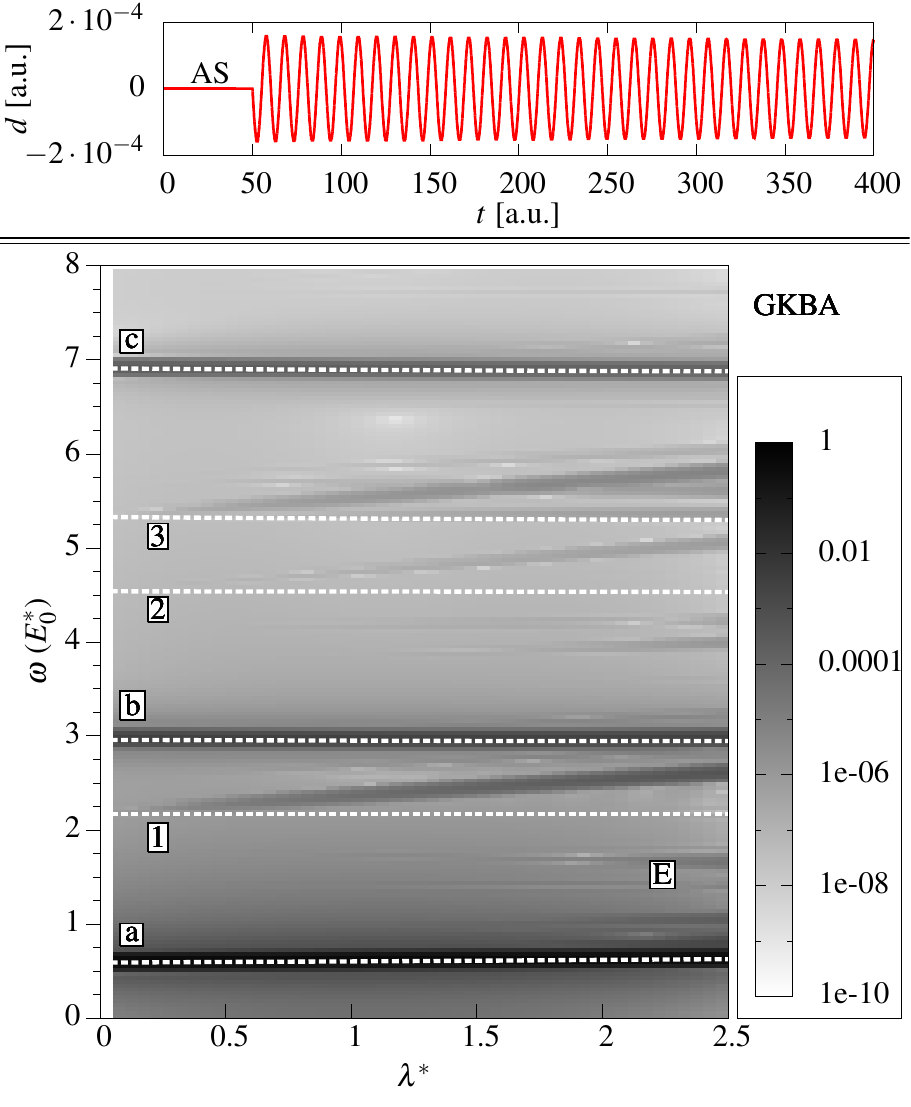}
\caption{Dipole excitation of a 2--electron quantum well. Top: Time--dependent dipole moment for coupling $\lambda^*=1$ for GKBA with adiabatic switching ("AS"). Bottom: Dipole excitation spectrum [Fourier transform of $d(t)$] for different coupling strengths $\lambda^*$. GKBA results with second Born self--energy (gray--scale) are compared to the results from exact diagonalization (dashed white lines). The letters (numbers) refer to one--(two--)electron excitations. Note, that the effective energy $E_0^*$ scales $\propto$ $\left(\lambda^*\right)^{-2}$, so the excitations nearly constant with $\lambda^*$ in this figure, effectively also scale $\propto$ $\left(\lambda^*\right)^{-2}$.}
\label{fig:Spectrum}
\end{figure}
\end{center} 
In Fig.~\ref{fig:Spectrum} the ground state dipole excitation spectrum of the 2--electron quantum well is presented for different values of the coupling parameter $\lambda$. In gray--scale the results from GKBA calculations using second order Born self--energy are shown. The white dashed lines represent the excitation energies from exact diagonalization (ED). Exemplarily for $\lambda^*=1$, the respective time--dependent dipole moment $d(t)$ can be seen in the figure above the spectrum. The first 50 a.u. of the propagation, where the dipole moment is zero, accounts for the adiabatic switching (denoted "AS" in the figure). While $d(t)$ appears to be monochromatic, in fact, it contains numerous additional frequencies which can only be resolved using a sufficiently long time propagation. The present GKBA calculation makes this possible. It has a total duration of $T=40000$ time steps and is readily performed within 24 hours for a few tens of basis functions.

Let us first discuss the general structure of the excitation spectrum that is obtained from the ED calculations. The excitations can be classified according to the number of electrons involved in the transitions, leading for the 2--electron quantum well to the distinction between single-- (SE) and double--excitations (DE). In particular, the double excitations are of high interest in the dynamics of correlated electronic systems. Obviously, they cannot be captured by mean--field type approaches such as Hartree--Fock.
For $\lambda^*=1$, the lowest excitation from the ground state, denoted by "a", is a SE of energy $\omega_a^{ED}=0.61\,E_0^*$. It is followed by a DE ("1"), with $\omega_1^{ED}=2.17\,E_0^*$, another SE ("b"), at $\omega_b^{ED}=2.95\,E_0^*$, two DEs, ("2") at $\omega_2^{ED}=4.54\,E_0^*$, respectively, ("3") at $\omega_3^{ED}=5.31\,E_0^*$ and another SE ("c") with energy $\omega_c^{ED}=6.89\,E_0^*$.

Consider now the results from the GKBA calculations. It can be clearly seen from Fig. \ref{fig:Spectrum} that it shows single excitations as well as double excitations, although the quality of their description differs significantly. While the SEs are very well described and are practically identical to the exact results over the whole range of coupling parameters $\lambda^*$, the double excitation energies only coincide in the limit $\lambda^* \rightarrow 0$, and the deviations from the exact result increase approximately linearly with $\lambda^*$, leading to a relative error of the order of 20 \% at $\lambda^*=2.5$ for the lowest DE ("1").
Thus the primary conclusion is that our NEGF approach within the GKBA is indeed able to reproduce the double excitations in the spectrum.
At the same time their energy appears with an incorrect $\lambda^*$--dependence in the present second Born approximation for the self--energy suggesting that not all processes leading to these excitations are captured.
For completeness, we note that, for coupling strengths $\lambda^* > 1$, additional excitations appear in the spectrum that cannot be attributed to real excitations in the system (cf., e.g. "E").
\begin{center}
\begin{figure}
\includegraphics[width=8.5cm]{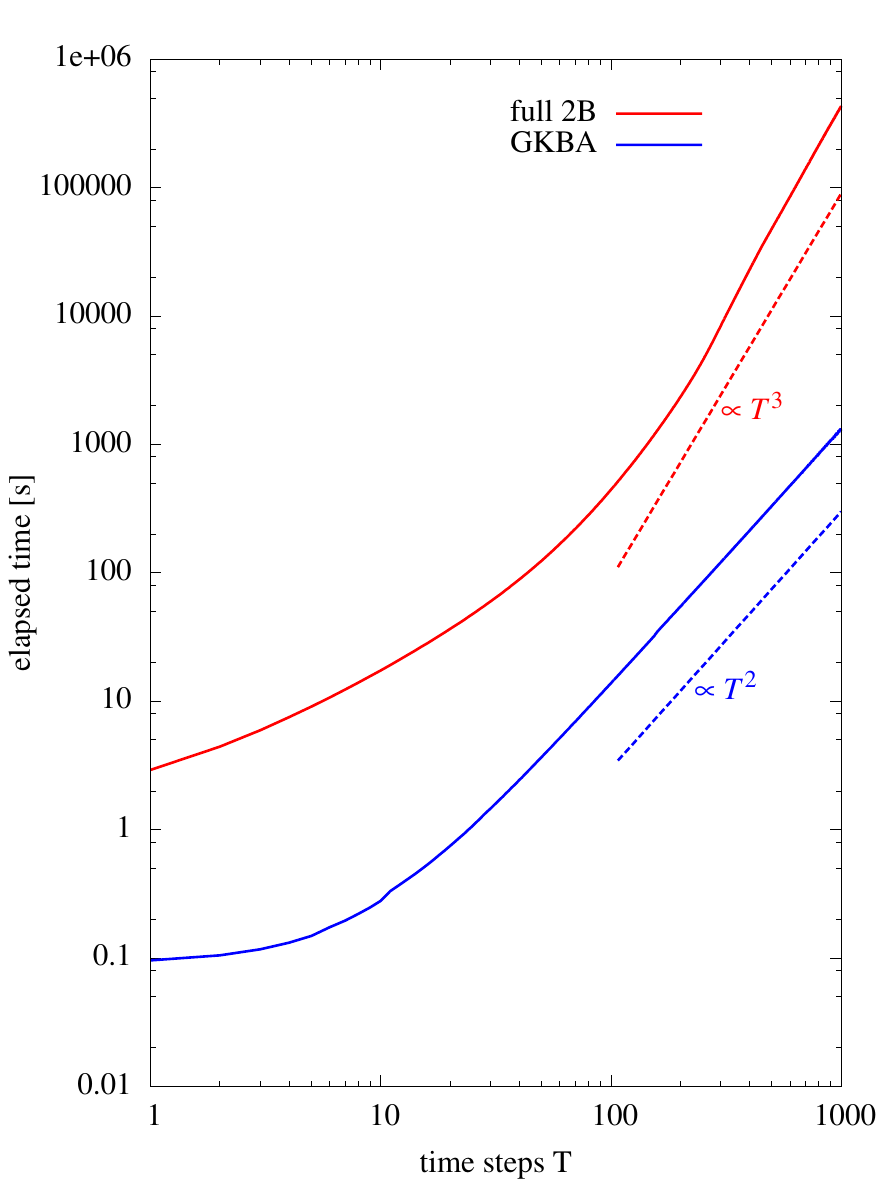}
\caption{Performance of second order Born NEGF simulations for a two--electron quantum well. GKBA lower (blue) lines vs. full two--time propagation, upper (red) lines. The total elapsed computation time is shown as a function of the number of propagation time steps $T$ for fixed step size $\delta t=0.01\,a.u.$. The dashed lines are guides to the eye indicating the expected asymptotic behaviour.}
\label{fig:perf}
\end{figure}
\end{center}

To analyze the numerical performance of the GKBA we tested it against full two--time calculations, both using the same second order Born self--energy. Fig. \ref{fig:perf} shows the scaling of the computation time with the propagation time $T$ on a single standard CPU.
From the graph it can be seen, that the full propagation scales with $T$ to a power of greater than $3$, we expect that it will converge to a scaling of $\propto T^3$ for longer time propagation. The GKBA, in contrast, scales only as $T^2$. This figure shows that the GKBA allows to increase the propagation duration $T$ by three or more orders of magnitude compared to two--time calculations.
This, in addition to the significant reduction of memory consumption, paves the way to much longer propagation times in the future.    
\section{Conclusions and outlook}
In this contribution we have shown recalled the idea and previous applications of the generalized Kadanoff-Baym ansatz and we demonstrate how to extend it to inhomogeneous, finite systems. A key for an efficient and consistent simulation was to correctly provide a correlated initial state via adiabatic switching of the interaction. For a first test of the accuracy of the approximations we have applied the formalism to a 2--electron quantum well model system and have studied the ground state dipole excitation spectrum for different coupling parameters. While our approach is easily applicable to systems containing more particles, the two-electron case allows for a benchmark against exact diagonalization results.

Our numerical results confirmed that the GKBA correctly recovers, besides single-particle excitations, also double excitations. These are presently of high interest for many applications in semiconductor optics and transport but cannot be obtained by standard tools such as time-dependent Hartree-Fock. At the same time, we have found that while the SE are reproduced with high accuracy, the DE are correctly captured only at small coupling. Inherent to the GKBA in second Born approximation is an incorrect coupling parameter dependence of the DE energy 
of the order $\propto \left(\lambda^*\right)^{-1}$, in contrast to the correct scaling of $\propto \left(\lambda^*\right)^{-2}$. 
This is in good agreement with our similar findings for the 4--electron quantum well case~\cite{KarstenQW}. Since a similar scaling is observed for the full two--time propagation within the second order Born self--energy we conclude, that this behavior of the DE energies is not a deficiency of the GKBA but indicates the limitations of the involved (weak coupling) second Born approximation for the self--energy. 
Evidently, higher terms in the Born series are required to restore the correct scaling. Therefore, in future work we will study higher order approximations for the selfenergy such as T--matrix-- or GW--approximation. It will be interesting to see whether the GKBA performs similarly well allowing again to omit the complicated integral terms in the full equations (\ref{eq:G_R_less}, \ref{eq:G_A_less}).

Our results based on the GKBA open up a broad variety of new many-body applications of inhomogeneous finite systems. In our recent work \cite{balzer_efficient_2010,balzer_time-dependent_2010} we demonstrated that these systems become tractable by using the FEDVR representation. Still there were essential limitations of full two-time calculations in terms of computation time and memory requirements.
These limitations can now be mitigated to a large extend with the help of the  GKBA by increasing the duration of the time propagation by more than three orders of magnitude. Not only does this allow for a much more accurate computation of electronic spectra, including double excitations. At the same time, true nonequilibrium problems such as nonlinear excitation and relaxation dynamics or pump-probe problems in inhomogeneous systems are now within reach of NEGF simulations.
\begin{acknowledgments}
This work was supported by the DFG via grant BO1366-9 and by computing time at the North-German Supercomputing Alliance (HLRN) via Grant No. shp0006.
\end{acknowledgments}


\begin{thebibliography}{10}

\bibitem{KadanoffBaym1962}
Kadanoff L and Baym G 1962 {\it {Q}uantum {S}tatistical {M}echanics} (W. A. Benjamin, Inc.: New York)

\bibitem{balzer_efficient_2010} Balzer K, Bauch S and Bonitz M 2010 {\it Phys. Rev.} A, \textbf{81} 022510

\bibitem{lipavsky_generalized_1986} Lipavský P, Spicka V and Velický B 1986 {\it Phys. Rev.} B \textbf{34} 6933

\bibitem{rios_time-dependent_2008} Rios A and Danielewicz P 2008 {\it AIPConf.Proc.} \b{995}, 98--103

\bibitem{KeldyshContour} Keldysh L 1964 {\it ZhETF} \b{47} 1515

\bibitem{martin_theory_1959} Martin P and Schwinger J 1959 {\it Physical Review} \b{115} 1342--1373

\bibitem{KeldyshKamenev} Kamenev A 2009 {\it{I}ntroduction to the {K}eldysh {F}ormalism}

\bibitem{LWR} Langreth D and Wilkins J 1972 {\it Phys. Rev. }B \textbf{6} 3189--3227

\bibitem{spicka_long_2005} Spicka V, Velický B and Kalvová A 2005 \newblock {\it Physica} E \textbf{29} 154--174

\bibitem{gartner_prb99} Gartner P, B\'anyai L and Haug H 1999 {\it Phys. Rev. }B \textbf{60} 14234--14241

\bibitem{gartner_prb06} Gartner P, Seebeck J and Jahnke F 2006 {\it Phys. Rev. }B \textbf{73} 115307

\bibitem{bonitz_jpcm96} Bonitz M, Kremp D, Scott D, Binder R, Kraeft W and K\"ohler H 1996 {\it J. Phys.: Cond. Matt.} \textbf{8} 6057

\bibitem{koehler_pre96} K\"ohler H 1996 {\it Phys. Rev. }E \textbf{53} 3145--3153

\bibitem{kwong_pss98} Kwong N, Bonitz M, Binder R, and K\"ohler H 1998 {\it Phys. Stat. Sol} B \textbf{206} 197-201

\bibitem{bonitz_epjb} Bonitz M, Semkat D and Haug H 1999 {\it Eur. Phys. J.} B \textbf{9} 309

\bibitem{HaugJauho} Haug H and Jauho A P 1996 {\it Quantum kinetics in transport and optics of semiconductors}, Springer Berlin

\bibitem{kremp_pre99} Kremp D, Bornath T, Bonitz M and Schlanges M 1999 {\it Phys. Rev. }E \textbf{60} 4725--4732

\bibitem{bonitz_cpp99} Bonitz M, Bornath T, Kremp D, Schlanges M and Kraeft W D 1999 {\it Contrib. Plasma Phys.} \textbf{39} 329

\bibitem{haberland_pre01} Haberland H, Bonitz M and Kremp D 2001 {\it Phys. Rev. }E \b{64} 026405

\bibitem{bonitz-book} Bonitz M 1998 {\it Quantum Kinetic Theory} Teubner, Stuttgart, Leipzig

\bibitem{bonitz_pla96} Bonitz M, and Kremp D 1996 {\it Phys. Lett.} A {\bf 212} 83 

\bibitem{velicky_jpcs06} Velický B, Kalvová A and Spicka V 2006 {\it J. Phys. Conf. Ser.} \textbf{35} 1--16

\bibitem{VetterWalecka} Fetter A L and Walecka J D 1971 {\it {Q}uantum {T}heory of {M}any--{P}article {Systems}}, {M}c{G}raw--{H}ill {B}ook {C}ompany, {N}ew {Y}ork

\bibitem{stan_time_2009} Stan A, Dahlen N E and van Leeuwen R 2009 {\it J. of Chem. Phys.} \textbf{130} 224101

\bibitem{born_beweis_1928} Born M and Fock V 1928 {\it Zeitschrift für Physik A Hadrons and Nuclei} \textbf{51} 165--180

\bibitem{kwong_prl00} Kwong N H and Bonitz M 2000 {\it Phys. Rev. Lett.} {\bf 84} 1768

\bibitem{balzer_time-dependent_2010} Balzer K, Bauch S and Bonitz M 2010 {\em Phys. Rev.} A \b{82} 033427

\bibitem{KarstenQW} Balzer K, Hermanns S and Bonitz M 2012 {\it Electronic double-excitations in quantum wells: solving the two-time
  {K}adanoff-{B}aym equations}, submitted to EPL
\end{thebibliography}
\end{document}